\documentclass[useAMS,usenatbib]{mn2e}

\usepackage{graphicx}

\title[A Phase Dependent Comparison of the Velocity Parameters of SiO v=1, J=1-0 and J=2-1 Maser Emission in Long Period Variables]{A Phase Dependent Comparison of the Velocity Parameters of SiO v=1, J=1-0 and J=2-1 Maser Emission in Long Period Variables}
\author[B. T. Indermuehle and G. C. McIntosh]{B. T. Indermuehle$^{1}$\thanks{E-mail:
balt.indermuehle@csiro.au (BTI)} and G. C. McIntosh$^{2}$\\
$^{1}$CSIRO Astronomy and Space Science, P O Box 76, Epping 1710 NSW, Australia\\
$^{2}$Division of Science and Mathematics, University of Minnesota, Morris, 600 East 4th Street, Morris, MN 56267, USA}
\begin{document}

\date{Accepted YYYY Month DD. Received 2013 August 9; in original form 2013 August 9}

\pagerange{\pageref{firstpage}--\pageref{lastpage}} \pubyear{2013}

\maketitle

\label{firstpage}

\begin{abstract}
We have examined the relationship between the velocity parameters of SiO masers and the phase of the long period variable stars (LPVs) from which the masers originate. The SiO spectra from the v=1, J=1-0 (43.122 GHz; hereafter $J_{1\rightarrow0}$) and the v=1, J=2-1 (86.2434 GHz; hereafter $J_{2\rightarrow1}$) transitions have been measured using the Mopra Telescope of the Australia Telescope National Facility.  One hundred twenty one sources have been observed including 47 LPVs contained in the American Association of Variable Star Observer Bulletin (2011). The epoch of maxima and the periods of the LPVs are well studied. This database of spectra allows for phase dependent comparisons and analysis not previously possible with such a large number of sources observed almost simultaneously in the two transitions over a time span of several years. The velocity centroids ($VCs$) and velocity ranges of emission ($VRs$) have been determined and compared for the two transitions as a function of phase. No obvious phase dependence has been determined for the $VC$ or $VR$. The results of this analysis are compared with past observations and existing SiO maser theory.
\end{abstract}

\begin{keywords}
Physical Data and Processes -- Masers. Stars --  Variables
\end{keywords}

\section{Background}
\subsection{$VCs$ and $VRs$}
The determination of the $VC$ and $VR$ were recently presented \citep{b5} and will only be briefly reviewed here.	Mathematically, the $VC$ is the sum of the antenna temperature $T_a$ in each velocity channel times the velocity with respect to the local standard of rest $v_{lsr}$ of the velocity channel over the range of emission divided by the sum of the $T_a$ in each velocity channel over the range of emission as shown in equation \ref{vc}.

\begin{equation}\label{vc}
VC=\frac{\sum{(T_a v_{lsr})}}{\sum{T_a}}
\end{equation}

The summation extends over the range of emission.

The $VR$ is calculated to be the region where the Ta exceeds three times the standard deviation of the antenna temperature of the background noise. The standard deviation is determined from velocity channels far away from the emission range of the source.

\citet{b5} examined the $VCs$ and $VRs$ of the SiO maser transitions without regard to the phase of the star. In this work we found that the $VR_{1\rightarrow0}$ was generally broader than the $VR_{2\rightarrow1}$, $6.4 km s^{-1}$ to $4.2 km s^{-1}$, respectively. The $VC_{1\rightarrow0}$ are slightly more positive than the $VC_{2\rightarrow1}$. These differences indicate that the $J_{1\rightarrow0}$ occurs in a dynamically different region of the circumstellar environment than the $J_{2\rightarrow1}$. This conclusion is consistent with the observational results of \citet{b10, b12}. The $VC$ is a well suited parameter to compare our single dish observations to VLBI observations, as it is weighted to the brightest emission. This effect is apparent in the figures B1-B3 in \citet{b24}. \citet{b25} have compared SiO and H$_2$O maser emission from a 401 evolved stars. They find that SiO emission is less dependent on the optical phase than the H$_2$O masers. While this is a relative conclusion only, it indicates no strong phase relationship for the $VR$, a finding which complements our own.

\citet{b9} concluded that $J_{1\rightarrow0}$ and $J_{2\rightarrow1}$ emission can form at comparable radii and in related columns of gas although their VLBI maps show few overlaps between the two transitions. Recent theoretical developments vary in predicting the radii at which the $J_{1\rightarrow0}$ and $J_{2\rightarrow1}$ emission occur. \citet{b2} have the $J_{2\rightarrow1}$ emission at larger or smaller radii than the $J_{1\rightarrow0}$ emission depending on the phase. \citet{b13} show no discernible difference in the radii at which the maser originate at any phase presented. 

\subsection{Phase Determination}
The phase of the star is the optical phase. It is determined by subtracting the Julian Day Number of the most recent stellar maximum from the Julian Day Number of the observation and dividing by the stellar period. The phase varies from zero to one. The stellar maxima and periods were obtained from the AAVSO Bulletin 74 \citep{b1}. Stellar periods were added or subtracted from the Bulletin 74 maxima as needed to determine the most recent maxima. \ref{tab:lpvs} gives the stellar periods used for the sources.

\subsection{Maser Velocity Parameters}
\citet{b2} investigated the dynamics of the circumstellar region in which the SiO masers originate and have provided the most thoroughly developed theory for the maser spectra in LPVs from phase 0.1 to 0.4. They model and depict a shock traveling out from the star generating different velocities, red shifted or more positive and blue shifted or more negative, at different distances from the star. As the shock travels out the velocities change as a function of distance from the star and phase. \citet{b2} predicted a phase dependent $VR_{1\rightarrow0}$ varying from about $8$ to $13 km s^{-1}$. The $VR_{2\rightarrow1}$ varies from about $7$ to $12 km s^{-1}$. The difference between the $VRs$ in the two transitions is phase dependent and difficult to quantify from the information presented, but the $VR_{1\rightarrow0}$ always appears to be greater than the $VR_{2\rightarrow1}$. In their Figure 12 they show that at a phase of 0.4 only the $J_{1\rightarrow0}$ emission should be present, at a phase of 0.3 the maximum $VR$ should occur in the $J_{1\rightarrow0}$ emission, several $km s^{-1}$ broader than the emission at phase of 0.1, and the $VC_{1\rightarrow0}$ should undergo a redward shift with increasing phase. No shift in the peak or $VC$ is indicated for the $J_{2\rightarrow1}$ transition.

\citet{b13} have developed a coupled escape probability model for SiO maser emission. They present several figures indicating the $VR$ in the $J_{1\rightarrow0}$ and $J_{2\rightarrow1}$ transitions. In their Figure 7, the $J_{2\rightarrow1}$ emission is consistently broader than the $J_{1\rightarrow0}$ emission, and the $VR_{1\rightarrow0}$ and $VR_{2\rightarrow1}$ vary by a factor of more than five over the stellar period. The emission at their epoch 6 is several times broader than the emission at the other epochs depicted. For their Epoch 11 the $J_{1\rightarrow0}$ is very weak and narrow.
Since different distances from the star are affected differently by the proposed shock travelling out from the star, it is reasonable to expect that masers forming at different distances from the star will exhibit different $VRs$ and slightly different $VCs$. The observations of the velocity parameters of the emission provide information on the locations as well as the motion of the masing material.

\section{Observations}
The 22 m Mopra radio telescope is located in New South Wales, Australia. At Mopra each maser source is first pointed on to ascertain adequate positioning inside the beam. The observations are then executed as 16 cycles on and 16 cycles off observation, thus lasting 64 seconds per on/off pair. For $J_{2\rightarrow1}$ observations the rms noise was about 1.8 Jy. For $J_{1\rightarrow0}$ observations the rms noise was about 0.6 Jy. The velocity resolution was $0.23 km s^{-1}$ ($J_{1\rightarrow0}$) and $0.12 km s^{-1}$ ($J_{2\rightarrow1}$).
The Mopra monitoring programme observes 121 sources since 2008 and is described in the data catalogue by \citet{b22}. LPVs, semi-regular variables, irregular variables, OH-IR stars, and the Orion SiO maser source were observed approximately monthly in $J_{1\rightarrow0}$ and $J_{2\rightarrow1}$ between 2008 and 2012.
LPVs were chosen for this analysis because they present a relatively uniform collection of sources and a group of stars for which theoretical models of the SiO maser emission have been developed. For the 47 LPVs observed, the stellar velocity, the number of observations in each transition, and the number of observations of both transitions within 24 hours are given in Table \ref{tab:lpvs}. The sources are listed in the AAVSO Bulletin \citep{b1} as LPVs with well determined maxima and periods. Non detections of the sources have not been included in the following analysis.

\section{Velocity Parameters and Results}
\subsection{$VC$ Comparison}
Since observations indicate and theories predict that the $J_{1\rightarrow0}$ and $J_{2\rightarrow1}$ masers exist at different distances from the star a phase dependent difference in the $VC_{2\rightarrow1}$ and $VC_{1\rightarrow0}$ might exist as the proposed shock propagates outward from the star. Figure 1 shows the $VC_{2\rightarrow1}$ – $VC_{1\rightarrow0}$ versus phase. No phase dependence of the $VC$ difference is obvious. There are also predictions that the $VC$ should have a redward shift in the $VC_{1\rightarrow0}$ versus phase \citep{b2}. Figure \ref{fig2} shows the $VC_{1\rightarrow0}$ at the phase indicated minus the $VC_{1\rightarrow0}$ at the earliest observed phase during a stellar cycle (normalized to the earliest observed phase). At least five observations of the source during an individual phase period had to be available to include the information in this analysis. This requirement insured that only data representing observations spread over the phase period were analyzed.  Again there is no obvious systematic redward or blueward shift . Figure \ref{fig3} presents the same information for the $J_{2\rightarrow1}$. For this transition the blueward spread is larger than the redward spread, but again no definitive trend is obvious.

\begin{figure}
  \includegraphics[width=1\columnwidth]{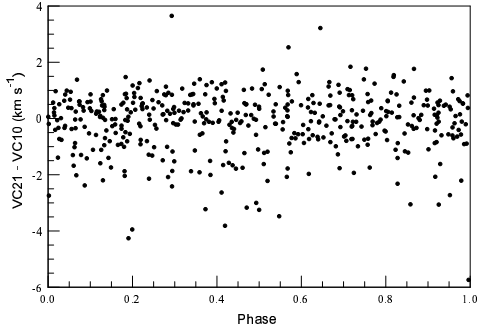}
  \caption{The difference between $VC_{2\rightarrow1}$ and $VC_{1\rightarrow0}$ versus Phase. 457 points are included in the plot.}
  \label{fig1}
\end{figure}

\begin{figure}
  \includegraphics[width=1\columnwidth]{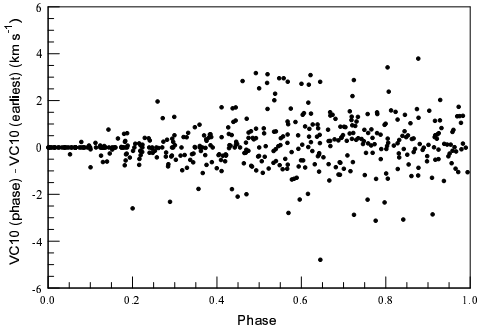}
  \caption{The change in $VC_{1\rightarrow0}$ versus phase. The $VC$ at the earliest phase observed has been subtracted from the $VCs$ at later phases. 442 points are included in the plot.}
  \label{fig2}  
\end{figure}

\begin{figure}
  \includegraphics[width=1\columnwidth]{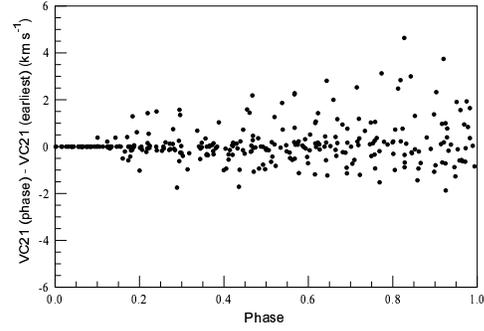}
  \caption{As in Figure \ref{fig2} for the $VC_{2\rightarrow1}$ data. 298 points are included in the plot.}
  \label{fig3}  
\end{figure}

\begin{figure}
  \includegraphics[width=1\columnwidth]{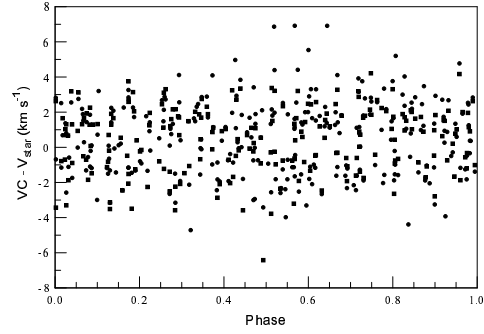}
  \caption{The difference between $VC$ -- $V_{star}$ versus Phase. The 277 $VC_{1\rightarrow0}$ points are indicated by circles. The 214 $VC_{2\rightarrow1}$ points are indicated by squares.}
  \label{fig4}  
\end{figure}

\citet{b20} subtracted the stellar velocity from the ``mean velocity'' of the $J_{1\rightarrow0}$ and the SiO v=2, J=1-0 emission from 43 late type stars. They concluded, based on the data shown in their Figure 5, ``The mean velocity of the 28 SiO v=1 and v=2 masers as a function of the optical phase shows that the redshifted emission is dominating during the phases from 0.3 to 0.8, while the blueshifted emission appears from phase $\approx 0.85$ and is dominating during phases 0.0-0.2 (or 1.0-1.2).'' Figure \ref{fig4} shows the results of a similar analysis with our data subtracting the stellar velocity from the velocity centroid. We have approximately 10 times as many data points as \citet{b20}. No phase dependent shift of the $VCs$ with respect to the stellar velocity is obvious.

\subsection{$VR$ Comparison}
\begin{figure}
  \includegraphics[width=1\columnwidth]{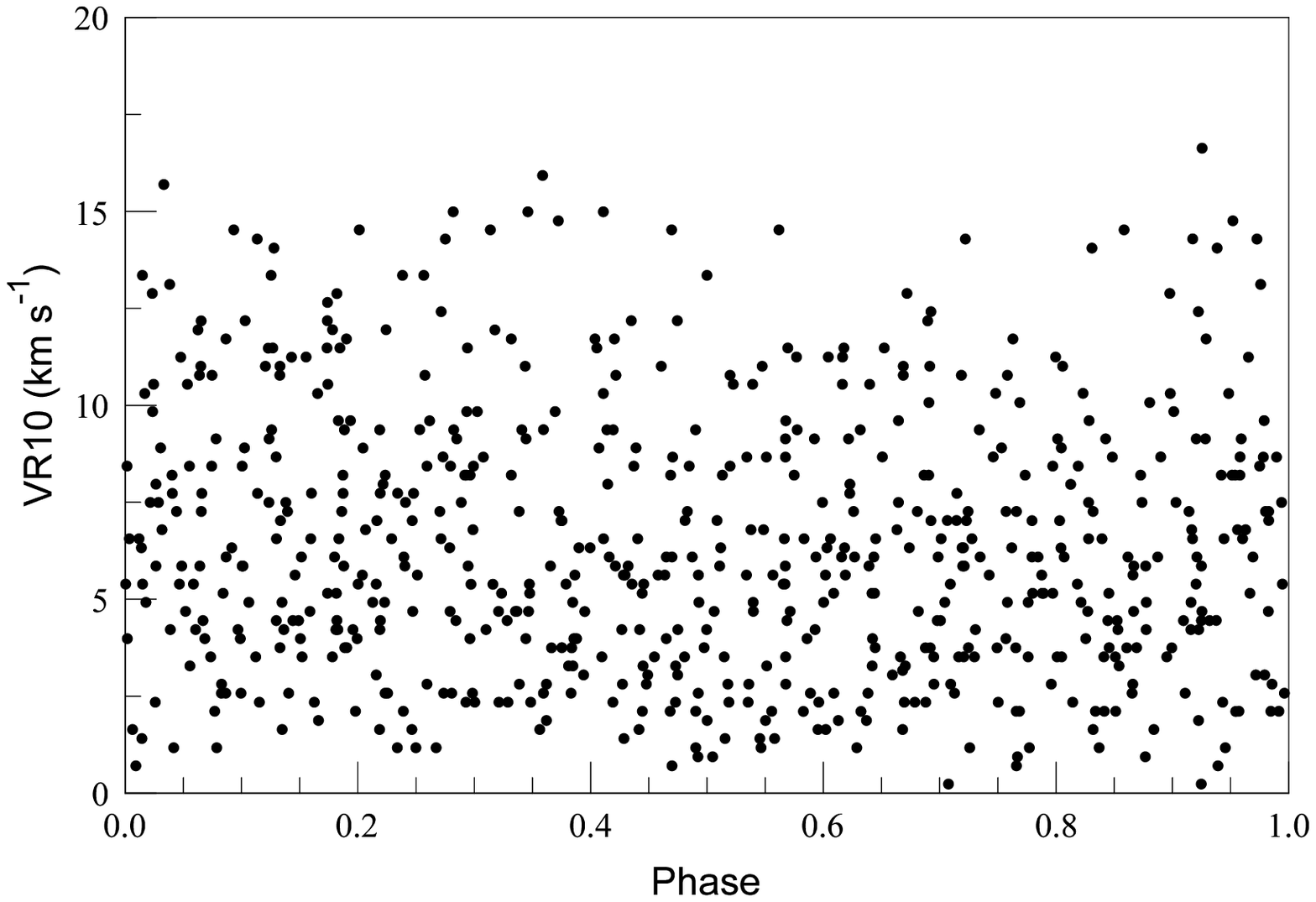}
  \caption{The $VR_{1\rightarrow0}$ versus the Phase. 635 points are included in the plot.}
  \label{fig5}  
\end{figure}

\begin{figure}
  \includegraphics[width=1\columnwidth]{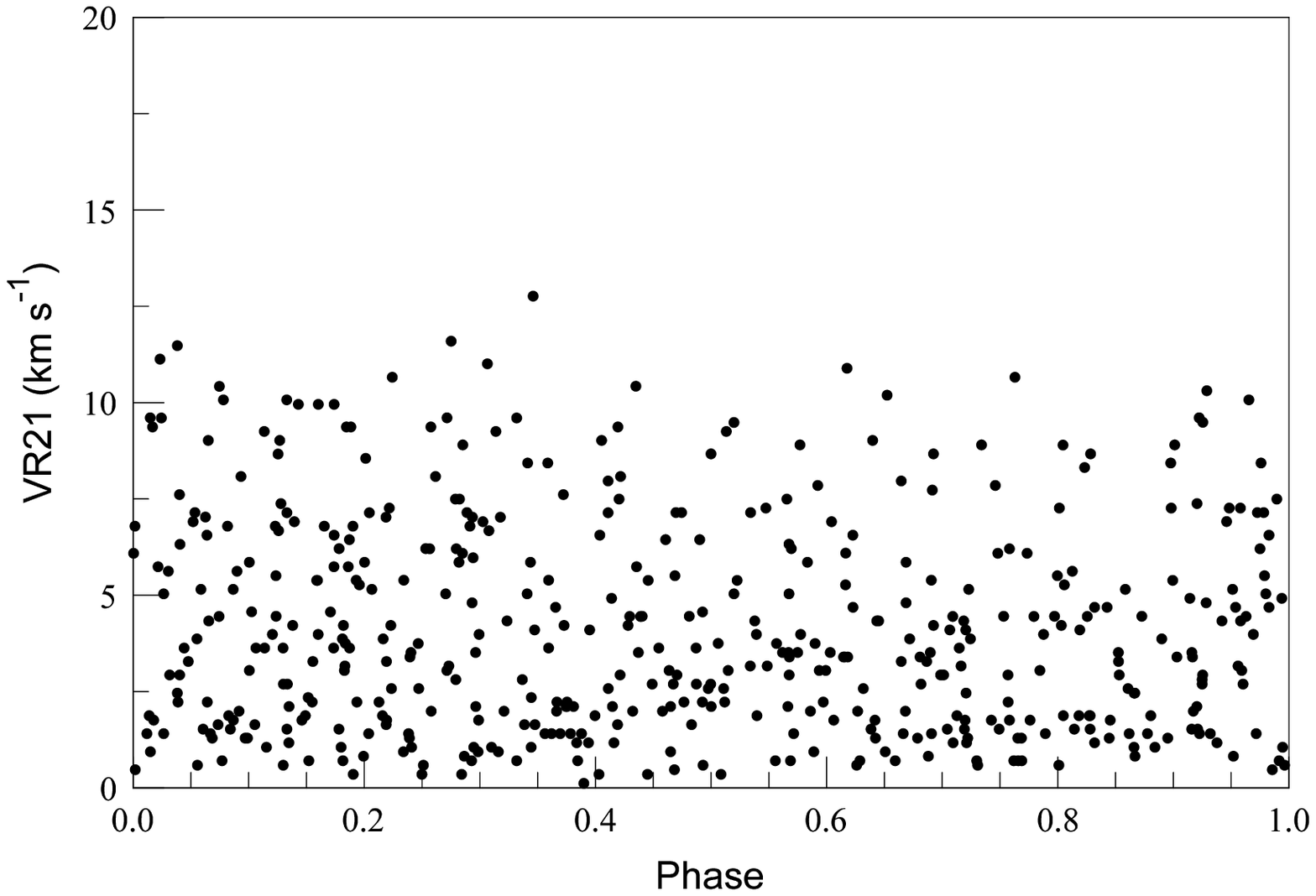}
  \caption{The $VR_{2\rightarrow1}$ versus the Phase. 485 points are included in the plot.}
  \label{fig6}  
\end{figure}

Many observers have concluded that the SiO maser integrated flux density is a maximum at a stellar phase of between 0.05 and 0.2 \citep{b3,b8}. \citet{b21} stated that there was a variable contrast of the integrated flux density over the period of the stars. The contrast was defined as the ``intensity ratio between consecutive maximum and minimum epochs.''  The average contrast was approximately three but ranged from four to 20 for Mira. These conclusions have generally been based on observations of the brightest maser sources and all the sources were not LPVs. 
While the integrated flux density versus phase has been well examined, the $VR$ has not been previously examined as a function of stellar phase. Figures \ref{fig5} and \ref{fig6} show the velocity ranges of the $J_{1\rightarrow0}$ and $J_{2\rightarrow1}$ transitions respectively. The velocity range of emission does not reproduce the phase dependence of the integrated intensity. There is no obvious maximum or minimum in the $VR$ of either transition versus phase. The $VR_{1\rightarrow0}$ generally exceeds the $VR_{2\rightarrow1}$ as stated in \citet{b5}. Figure 8 in \citet{b25} shows the full width at zero power versus optical phase for SiO masers, which is a similar metric as we are employing and showing in our Figure \ref{fig5}. It is interesting to see that they do not find a strong correlation either between optical phase and peak emission from SiO masers in their large sample of 401 late type stars.

\section{Future Work}
In the future the Mopra SiO maser database will be used to look for periodicities in velocity parameters of individual sources. L2 Puppis, a semi-regular variable, has been examined and a 139 day periodicity in the SiO $VC$ has been found \citep{b6}. This periodicity likely indicates a non-radial pulsation of the star. Periodicities in the velocity parameters of the star could indicate various interesting phenomena – planets, non-radial effects, or other asymmetries in the circumstellar environment.  The case of L2 Puppis is similar to that of findings by a VLBI study of TX Cam (a source too far north to be in the Mopra catalogue) in which periodicity in SiO emission has been observed by \citet{b23}.

\section{Conclusions}
The Mopra database provides the first large data set of LPV SiO maser spectra (essentially simultaneous observations in $J_{2\rightarrow1}$ and $J_{1\rightarrow0}$ from 2008 until 2012) to allow the comparison of the $VCs$ and $VRs$ versus phase with theoretical model predictions. The velocity comparisons extracted from these observations will inform and constrain the development of future models of the circumstellar environment and maser dynamics. The analysis of the $VCs$ and $VRs$ of 47 LPVs as a function of phase shows no shifts or variations that indicate the passage of a shock through the circumstellar material.

\section*{Acknowledgments}
We would like to acknowledge the efforts of Remi Patriot, Chad Reverman, and Emma Molden in the data reduction that forms the basis for this paper and the support of the University of Minnesota, Morris, in this work.
The Mopra radio telescope is part of the Australia Telescope National Facility which is funded by the Commonwealth of Australia for operation as a National Facility managed by CSIRO. The University of New South Wales Digital Filter Bank used for the observations with the Mopra Telescope was provided with support from the Australian Research Council.

\begin{table*}
 \centering
 \begin{minipage}{170mm}
  \caption{Sources, number of observations, and source information. Details of the individual observations and spectra can be found at http://www.narrabri.atnf.csiro.au/cgi-bin/obstools/siomaserdb.cgi}
\label{tab:lpvs}  
  \begin{tabular}{lcccccrr}
  \hline
  Source & Epoch of First & J10 & J21 & Peak $v_{j=1-0}$ & Peak $v_{j=2-1}$ & $v_{star}$ & Stellar\\
         & J10 or         & Observations & Observations & [$km s^{-1}$] & [$km s^{-1}$] & [$km s^{-1}$] & Period\\  
         & Simultaneous &   &   &  &  & (see \footnote{\label{fb15}If single data, or data in bracket, source is \citet{b15}. Default source is \citet{b14}}) & [Days]\\
         & Observation & & & & & & (see \footnote{Default source unless otherwise noted is \citet{b14}})\\
  \hline
  
Z Peg & 2010 Oct 21 & 6 & 1 & -27.7 $\pm$ 0.9 & -26.4 $\pm$ 0.0 & -31.5 $\pm$ 5 & 334.8 \\
S Scl & 2009 Apr 15 & 21 & 18 & 6.9 $\pm$ 1.9 & 7.7 $\pm$ 1.1 & 35 $\pm$ 5 & 362.57 \\
R Psc & 2011 Jan 12 & 4 & 0 & -58.2 $\pm$ 0.3 & - & -45 $\pm$ 10  & 344.5  \\
Mira & 2009 Oct 5 & 22 & 24 & 48.7 $\pm$ 1.7 & 48.1 $\pm$ 1.0 & 46.3 [63.8 $\pm$ 0.9] \footnote{\label{fb16}In brackets, the data from \citet{b16}} & 331.96 \\
R Cet & 2010 Jan 19 & 13 & 0 & 31.9 $\pm$ 1.4 & - & 42 $\pm$ 5  & 166.24 \\
R Hor & 2009 Oct 5 & 22 & 19 & 38.2 $\pm$ 1.7 & 36.1 $\pm$ 1.7 &  60 $\pm$ 5 & 407.6 \\
W Eri & 2010 Mar 16 & 21 & 12 & -1.0 $\pm$ 1.1 & -0.4 $\pm$ 1.1 & 26 $\pm$ 5  & 376.63 \\
R Tau & 2011 Aug 13 & 18 & 7 & 15.4 $\pm$ 1.4 & 16.3 $\pm$ 0.9 & 13.5 [32.5 $\pm$ 5 ] \ref{fb16} & 320.9 \\
R Cae & 2009 Apr 21 & 20 & 17 & -0.2 $\pm$ 1.0 & -0.2 $\pm$ 1.2 & 14 \footnote{\label{fb17}Data from \citet{b17}}  & 390.95 \\
T Lep & 2009 Oct 5 & 20 & 14 & -28.1 $\pm$ 1.6 & -28.3 $\pm$ 0.6 & -4 $\pm$ 5   & 368.13 \\
S Pic & 2009 Apr 22 & 22 & 25 & -3.2 $\pm$ 1.2 & -2.4 $\pm$ 0.5 & 26 \ref{fb17}  & 428 \\
R Oct & 2009 Oct 5 & 20 & 19 & 20.5 $\pm$ 1.4 & 19.2 $\pm$ 1.2 & 46 $\pm$ 5  & 405.39 \\
S Ori & 2009 Apr 21 & 24 & 20 & 12.3 $\pm$ 1.9 & 12.0 $\pm$ 1.3 & 22 $\pm$ 5 & 414.3 \\
S Col & 2010 Jan 19 & 21 & 19 & 62.2 $\pm$ 0.8 & 63.0 $\pm$ 0.5 & 73 $\pm$ 5  & 325.85 \\
U Ori & 2009 Oct 5 & 24 & 25 & -37.9 $\pm$ 1.5 & -38.1 $\pm$ 0.3 & -39.0 [-20.8 $\pm$ 0.9 ] \ref{fb15} & 368.3 \\
R Cnc & 2009 Apr 28 & 20 & 24 & 14.1 $\pm$ 2.8 & 14.2 $\pm$ 2.1 & 35.42 $\pm$ 0.52 \footnote{\label{fb18}Data from \citet{b18}}  & 361.6 \\
W Cnc & 2009 Apr 28 & 10 & 2 & 36.2 $\pm$ 1.6 & 35.1 $\pm$ 1.8 & 49 $\pm$ 5  & 393.22 \\
R Car & 2009 Oct 5 & 16 & 14 & 7.4 $\pm$ 0.8 & 8.1 $\pm$ 1.2 & 28.1 $\pm$ 0.9  & 308.71 \\
X Hya & 2009 Apr 28 & 19 & 11 & 26.7 $\pm$ 1.0 & 26.7 $\pm$ 1.0 & 42 $\pm$ 10  & 301.1 \\
R LMi & 2009 Oct 5 & 18 & 16 & 0.3 $\pm$ 1.9 & 0.8 $\pm$ 1.1 &  1.0 [10 $\pm$ 5] \ref{fb15} & 372.19 \\
R Leo & 2009 Apr 28 & 20 & 26 & 0.0 $\pm$ 1.7 & -1.0 $\pm$ 1.2 & 0.5 [13.4 $\pm$ 0.9] \ref{fb15}  & 309.95 \\
W Leo & 2011 Dec 1 & 3 & 1 & 46.6 $\pm$ 2.7 & 46.3 $\pm$ 0.0 & 54 $\pm$ 2  & 391.75 \\
X Cen & 2010 Jan 19 & 19 & 11 & 28.0 $\pm$ 1.0 & 28.7 $\pm$ 0.7 & 38 $\pm$ 2  & 315.2 \\
T Vir & 2011 Oct 14 & 3 & 0 & 5.5 $\pm$ 0.4 & - & 22 $\pm$ 5  & 339.47 \\
R Hya & 2009 Apr 28 & 22 & 23 & -9.6 $\pm$ 1.2 & -9.5 $\pm$ 1.2 &  -10.5 [-10.4 $\pm$ 0.9] \ref{fb15} & 385 [388.87] \\
S Vir & 2009 Apr 28 & 22 & 13 & 10.5 $\pm$ 1.9 & 10.2 $\pm$ 1.2 & 10 $\pm$ 2  & 375.1 \\
RU Hya & 2011 Aug 15 & 21 & 6 & -3.3 $\pm$ 1.3 & 4.4 $\pm$ 1.6 & 2 $\pm$ 5  & 331.5 \\
RS Vir & 2010 Apr 12 & 19 & 9 & -12.0 $\pm$ 0.7 & -12.5 $\pm$ 1.0 & -10.0 [-26 $\pm$ 5] \ref{fb15} & 353.95 \\
S Ser & 2010 May 27 & 18 & 3 & 22.0 $\pm$ 2.1 & 21.5 $\pm$ 0.9 & 23.0 [12 $\pm$ 0.95] \ref{fb15} & 371.84 \\
RS Lib & 2009 Oct 6 & 21 & 13 & 7.2 $\pm$ 1.6 & 6.6 $\pm$ 1.9 & 10.0 [-5 $\pm$ 5] \ref{fb15} & 217.65 \\
R Ser & 2009 Apr 28 & 17 & 6 & 31.6 $\pm$ 2.4 & 33.2 $\pm$ 0.6 & 32.0 [23.7 $\pm$ 2] \ref{fb15} & 356.41 \\
RU Her & 2010 Apr 12 & 17 & 11 & -11.0 $\pm$ 3.0 & -11.4 $\pm$ 1.3 & -25 $\pm$ 2  & 484.83 \\
U Her & 2009 Oct 5 & 19 & 21 & -16.2 $\pm$ 0.2 & -15.8 $\pm$ 0.4 &  -15.0 [-27.6 $\pm$ 2] & 406.1 \\
T Oph & 2010 Apr 12 & 13 & 13 & -31.6 $\pm$ 0.8 & -31.5 $\pm$ 0.7 & -46 \footnote{\label{fb19}Data from \citet{b19}}  & 366.82 \\
X Oph & 2009 Oct 5 & 21 & 22 & -57.5 $\pm$ 1.3 & -58.0 $\pm$ 1.3 & -71 $\pm$ 10   & 328.85 \\
R Aql & 2009 Oct 5 & 18 & 18 & 48.3 $\pm$ 1.3 & 48.3 $\pm$ 2.0 & 48.5 [32 $\pm$ 0.9] \ref{fb15} & 279 [284.2] \\
RT Aql & 2009 Oct 5 & 17 & 9 & -30.6 $\pm$ 1.6 & -30.9 $\pm$ 0.7 & -30.5 [-41.0 $\pm$ 2] \ref{fb15} & 327.11 \\
S Pav & 2009 Oct 5 & 20 & 16 & -19.0 $\pm$ 1.6 & -19.4 $\pm$ 2.0 & -22.0 $\pm$ 2  & 380.86 \\
RR Sgr & 2010 Mar 15 & 15 & 9 & 97.7 $\pm$ 2.5 & 96.2 $\pm$ 2 & 85 $\pm$ 5  & 336.33 \\
RR Aql & 2009 Oct 5 & 18 & 16 & 31.5 $\pm$ 3.1 & 28.9 $\pm$ 2.1 &  27.5 [11.0 $\pm$ 2] & 394.78 \\
RU Cap & 2010 Apr 12 & 5 & 3 & 6.5 $\pm$ 0.1 & 6.7 $\pm$ 0.1 & -3 \ref{fb18}  & 347.37 [347.73] \\
W Aqr & 2009 Oct 5 & 21 & 15 & 0.5 $\pm$ 1.2 & 1.0 $\pm$ 0.8 & -0.8 [-15 $\pm$ 5] \ref{fb15} & 381.1 [376.1] \\
V Peg & 2011 Oct 12 & 1 & 0 & -18.8 $\pm$ 0.0 & - & -25 $\pm$ 5  & 302.35 \\
S Gru & 2009 Oct 5 & 9 & 10 & -24.1 $\pm$ 3.9 & -27.5 $\pm$ 0.7 & -21 \ref{fb18}  & 401.51 \\
R Peg & 2009 Apr 20 & 23 & 18 & 24.7 $\pm$ 1.6 & 24.0 $\pm$ 1.6 & 24.5 [20 $\pm$ 5] \ref{fb15} & 378.1 \\
W Peg & 2009 Apr 20 & 23 & 18 & -16.7 $\pm$ 0.5 & -16.1 $\pm$ 0.7 & -21 $\pm$ 5  & 345.5 \\
R Aqr & 2009 Apr 20 & 24 & 26 & -22.5 $\pm$ 1.8 & -22.7 $\pm$ 1.6 & -25.4 [-22.0 $\pm$ 0.9] \ref{fb15} & 386.96 [390.0] \\

\hline
\end{tabular}
\end{minipage}
\end{table*}

\appendix

\label{lastpage}

\end{document}